\documentclass[journal]{IEEEtran}
\usepackage{amsfonts}
\usepackage{amssymb}
\usepackage{amsmath}
\usepackage{diagbox}
\usepackage{algorithm}
\usepackage{multirow}
\usepackage{xcolor}
\usepackage{graphicx}
\usepackage{cite}
\usepackage{exscale}
\usepackage{relsize}
\usepackage{algpseudocode}
\usepackage{graphics}
\usepackage{mathrsfs}
\usepackage{siunitx}
\usepackage{threeparttable}
\usepackage{url}
\usepackage{booktabs}
\usepackage{lipsum}
\usepackage{array}

\newcommand{\bm}[1]{\mbox{\boldmath{$#1$}}}

\newtheorem{Theo}{Theorem}
\newtheorem{Prop}{Proposition}
\newtheorem{Lem}{Lemma}

\ifCLASSOPTIONcompsoc
\usepackage[caption=false,font=normalsize,labelfont=sf,textfont=sf]{subfig}
\else
\usepackage[caption=false,font=footnotesize]{subfig}
\fi

\usepackage[T1]{fontenc}

\title{Random Matrix based Physical Layer Secret Key Generation in Static Channels}

\author{Zhuangkun Wei\textsuperscript{1}, Weisi Guo\textsuperscript{1,2}

\thanks{\textsuperscript{1}School of Aerospace, Transport, and Manufacturing, Cranfield University, UK. 
\textsuperscript{2}The Alan Turing Institute, UK.}}

\begin{document}

\maketitle

\begin{abstract}
Physical layer secret key generation exploits the reciprocal channel randomness for key generation and has proven to be an effective addition security layer in wireless communications. However, static or scarcely random channels require artificially induced dynamics to improve the secrecy performance, e.g., using intelligent reflecting surface (IRS). One key challenge is that the induced random phase from IRS is also reflected in the direction to eavesdroppers (Eve). This leakage enables Eve nodes to estimate the legitimate channels and secret key via a globally known pilot sequence.
To mitigate the secret key leakage issue, we propose to exploit random matrix theory to inform the design of a new physical layer secret key generation (PL-SKG) algorithm. We prove that, when sending appropriate random Gaussian matrices, the singular values of Alice's and Bob's received signals follow a similar probability distribution. Leveraging these common singular values, we propose a random Gaussian matrix based PL-SKG (RGM PL-SKG), which avoids the usages of the globally known pilot and thereby prevents the aforementioned leakage issue.
Our results show the following: (i) high noise resistance which leads to superior secret key rate (SKR) improvement (up to 300\%) in low SNR regime, and (ii) general improved SKR performance against multiple colluded Eves. We believe our combination of random matrix theory and PL-SKG shows a new paradigm to secure the wireless communication channels.
\end{abstract}


\section{Introduction}
Wireless communications are vulnerable and easily to be eavesdropped due to their broadcasting property. Traditional cryptography suffers from the high latency and unguaranteed secrecy, given their computational complexity based key management and distribution \cite{8883127}, which therefore makes them less attractive to secure the confidentiality in swift and highly mobile wireless communications. To protect the wireless communications, physical layer security (PLS) has been proposed and attracted a wide-range of both academic and industrial attentions \cite{bloch2011physical,6739367,7120011,5422766,8883128,8758230,8758975,8403278}. The essential idea is to exploit the dynamics and random fading of wireless channels, which avoids the usages of complex and high latency cryptography techniques, and thereby renders a promising prospective in securing the wireless communications. 

\subsection{Literature Review}
From academic point of view, PLS techniques can be categorized as two families.

\subsubsection{Key-Less PLS}
The first one is key-less, and exploits the superiority of legitimate channels over wiretap channels. To maximize the secrecy rate, power and trajectory optimizations \cite{8618602,8643815}, beamforming (jamming) \cite{7059241,8533374,6646279,8438310}, artificial noise \cite{8438310,9201173,6094170}, and cooperation \cite{7944621,7004893} methods have been designed, which are able to protect the communication confidentiality under the awareness of eavesdroppers (Eve) channel state information (CSI). One difficulty lies in the unknown of the Eves (e.g., in silent mode). In such a case, the above methods may either become impractical or require other high complexity based technologies (e.g., camera or radar \cite{7807176,6888953}) to search and estimate the exact locations of Eves. Other key-less methods, e.g., robust (worst case) optimization \cite{8392472} and intercept probability security region \cite{8456560}, are not always feasible with unknown Eves' CSIs. 

\subsubsection{Physical Layer Secret Key Generation}
The second family is referred to as physical layer secret key generation (PL-SKG) \cite{7393435,6739367}. The idea is to exploit the spatial \& temporal -correlated randomness and dynamic fading of the legitimate channels to generate secret key \cite{7393435,5422766,4036441,5371757}. This is able to provide a promising secret key rate (SKR) if Eves are located farther than half of the wavelength \cite{9000831}. Leveraging this idea, key generation schemes using legitimate CSIs, received signal strength (RSS), and channel impulse response (CIR) have been well-studied \cite{7809064,5999759,azimi2007robust,mathur2008radio,jana2009effectiveness}. The researches are also extended to secure the relay communications \cite{7299681}, and communication networks \cite{8352789}. These studies all have good performance in rich randomness scenarios. However, when it comes to the scarce randomness environment, the legitimate channel properties for key generation will be static (quasi) over a long period, which unfortunately enables the Eves to intercept and estimate and thereby weakens the communication security. 

To address the static channel scenarios, recent advances on intelligent reflecting surface (IRS) have been regarded as a promising technologies to generate randomness and dynamics of fading via its phase controller \cite{9361290,9442833,staat2020intelligent,9360860}. To be specific, by randomly selected an IRS phase and remains in each coherent time, the spatial \& temporal -correlated randomness of legitimate channels can be created and exploited for SKG. Leveraging this, the work in \cite{9442833} randomly selects IRS phases from a discrete set, and provides a theoretical high SKR in the face of a single Eve. The work in \cite{9298937} further designs a selection of random IRS phase, by maximizing the SKR against multiple non-colluded Eves. However, none of them considers the colluded Eve cases. In essence, the randomness induced by the IRS phases is not only correlated on the legitimate nodes, but can also be revealed in the received signals of Eves. In this view, if Eves collude in a way and estimate the IRS phase in one coherent time, may the SKR decreases and the current PL-SKG methods become vulnerable. This thereby constitutes the motivations of this work.

\subsection{Contributions}
In this work, we combine the random matrix theory into the PL-SKG process, and uncover an unique property for the singular value attributes of the reciprocal legitimate channels.
To be specific, a random Gaussian matrix based PL-SKG (RGM PL-SKG) is proposed. The main advantage lies in the usages of the unknown random Gaussian signals, instead of the globally known pilot sequences, to derive the reciprocal channel properties for PL-SKG. This thereby avoids the Eves to estimate the IRS-induced randomness, since the random Gaussian signals are unknown to all users. The main contributions are given as follows:

(1) In IRS-assisted static channels, we provide a case of colluded Eves, which is able to use the globally known pilot sequences to estimate the IRS random phase for each coherent time. In this view, the colluded Eves are able to compute the reciprocal legitimate channel properties and subsequently derive the secret key. 

(2) To address the secret key leakage, we propose the RGM PL-SKG, which uses the unknown random Gaussian matrices instead of the global pilot sequences to obtain the reciprocal channel properties for SKG. Here, we prove that, by sending Gaussian random matrices, the singular values of Alice's and Bob's received signal matrices follow a similar probability distribution (see Theorem \ref{theo2} and Proposition \ref{prop1}). As such, by using the singular values of the received signals, Alice and Bob are able to generate the secret key, which cannot be estimated and obtained by Eves, given the sending of unknown random Gaussian matrices. 

(3) We evaluate our proposed RGM PL-SKG in IRS-assisted millimeter-wave communications. The results firstly show the outstanding and stable SKR performance in low SNR region, due to the noise resistance ability of the received signals' singular values. Then, compared to current global pilot based PL-SKG methods, our RGM PL-SKG illustrates a high SKR when combating the colluded Eves. The results therefore validate our proposed scheme, which gives a novel idea to design the PL secret key to secure the wireless communications.

The rest of this paper is structured as follows. In Section II, we provide the millimeter-wave based IRS-assisted communication model. Also, the colluded Eves are provided. In Section III, we elaborate our proposed RGM PL-SKG, and analyze its SKR. In Section IV, we show the simulation results. We finally conclude this paper in Section V.

\section{System Model and Problem Formulation}

\subsection{IRS-assisted Communication Model}
\begin{figure*}[!t]
\centering
\includegraphics[width=6.5in]{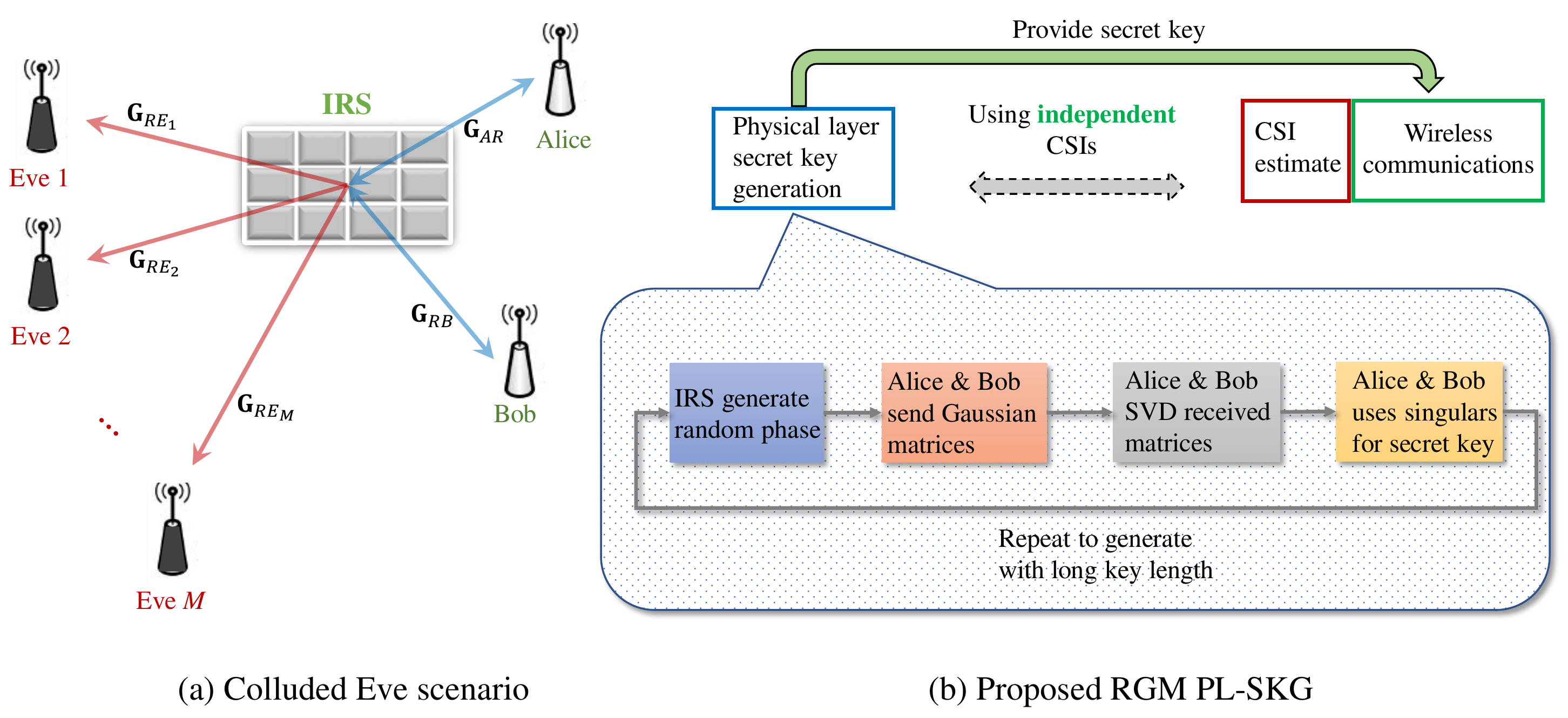}
\caption{IRS-assisted scarce randomness (static) wireless communications with (a) colluded Eves and (b) the designed RGM PL-SKG.}
\label{figm1}
\end{figure*}

In this work, we consider an IRS-assisted mmWave communication system, which includes a pair of legitimate users (Alice and Bob), $M\in\mathbb{N}^+$ Eves, and an IRS to secure the communication between legitimate users, illustrated in Fig. \ref{figm1}(a). Different from the key-less PLS that uses IRS to enhance the legitimate channels over wiretap ones \cite{9428001,8742603,9133130,9201173}, the IRS here is to generate randomness by randomly shifting its phase controller, for secret key generations between legitimate nodes \cite{9361290,staat2020intelligent,9360860,9442833}. 

The legitimate users (Alice and Bob) and all Eves are assumed to be uniform linear array (ULA) with the numbers of antennas as $N_A$, $N_B$ and $N_E$ respectively. The IRS is an uniform planar array (UPA) with $N_R=N_{R,x}\times N_{R,y}$ reflecting elements. As such, the channels between Alice and Bob, and from Alice to Eves can be specified as \cite{9103231,8811733}:
\begin{equation}
\label{eq1}
\begin{aligned}
    &\mathbf{H}_{AB}=\mathbf{G}_{RB}\cdot diag\left(\mathbf{w}\right)\cdot\mathbf{G}_{AR}+\mathbf{G}_{AB},\\
    &\mathbf{H}_{BA}=\mathbf{H}_{AB}^H,\\
    &\mathbf{H}_{AE_m}=\mathbf{G}_{RE_m}\cdot diag(\mathbf{w})\cdot\mathbf{G}_{AR}+\mathbf{G}_{AE_m},
\end{aligned}
\end{equation}
In Eq. (\ref{eq1}), $\mathbf{H}_{AB}$ represents the channel from Alice to Bob, and $\mathbf{H}_{BA}$ is the reciprocal channel from Bob to Alice. $\mathbf{H}_{AE_m}$ is the channel from Alice to Eve $m$ ($m\in\{1,\cdots,M\}$). $\mathbf{w}\triangleq[\upsilon_1\exp(j\theta_1),\cdots,\upsilon_{N_R}\exp(j\theta_{N_R})]^T$ is the IRS phase controller vector \cite{8811733}. In this work, we assign $\upsilon_1=\cdots=\upsilon_{N_R}=1$, and randomly select the phase $\theta_i,i\in\{1,\cdots,N_R\}$ over $[0,2\pi]$. 
$\mathbf{G}_{ab}\in\mathbb{C}^{N_a\times N_b}$, $a\neq b\in\{A,R,B,E_1,\cdots,E_M\}$ represents the direct channel from $a$ to $b$, which is assumed to be static over a long period. 

We then formulate these direct channels via the narrow-band geometric channel model, i.e., \cite{6847111}
\begin{equation}
\begin{aligned}
\label{direct_channels}
    \mathbf{G}_{ab}&=\sqrt{\frac{N_a\cdot N_b}{\rho_{ab}}}\cdot\sum_{l=1}^{L_{ab}} g_{ab,l}\cdot\mathbf{f}_b(\alpha_{ab,l},\beta_{ab,l})\cdot\mathbf{f}_a^H(\theta_{ab, l},\gamma_{ab,l}),\\ &~~~~~~~~~~~~~~~~~~~~~~~~~~a\neq b\in\{A,R,B,E_1,\cdots,E_M\}
\end{aligned}
\end{equation}
In Eq. (\ref{direct_channels}), the subscript $a$, and $b$ indicates the direct channel from $a$ to $b$. $\rho_{ab}$ represents its average path-loss. $L_{ab}$ denotes the number of path. For each path $l\in\{1,\cdots,L_{ab}\}$, $g_{ab,l}$ is the complex gain, $\alpha_{ab,l}$ ($\beta_{ab,l}$) is the azimuth (elevation) angle of arrival (AoA), and $\theta_{ab,l}$($\gamma_{ab,l}$) is the azimuth (elevation) angle of departure (AoD). The functions $\mathbf{f}_a(\cdot,\cdot)$ and $\mathbf{f}_b(\cdot,\cdot)$ are to generate the channel response given the AoA and AoD, and have a general form as \cite{6847111,8383706}:
\begin{equation}
\label{response}
\begin{aligned}
    \mathbf{f}(\alpha,\beta)=&\left[1~\exp(ju)~\cdots~\exp(j(x-1)u\right]^T\\
    &\otimes[1~\exp(jv)~\cdots~\exp(j(y-1)v]^T,\\
    u\triangleq&2\pi d\cos(\beta)/\xi.~v\triangleq2\pi d\sin(\beta)\sin(\alpha)/\lambda
\end{aligned}
\end{equation}
where $\otimes$ is the Kronecker product operator, $\lambda$ is the wavelength, and $d$ is the antenna spacing (usually assigned as $d=\lambda/2$). For the UPA IRS, we have $x=N_{R,x}$ and $y=N_{R_y}$. For the ULA Alice (Bob) $\mathbf{f}(\alpha,\beta)$ degenerate into $\mathbf{f}(\alpha,\pi/2)$ with $x=1$, $y=N_A$($N_B$).

\subsection{Globally Known Pilot based PL-SKG}
Given the IRS-assisted communication model in Eq. (\ref{eq1}), PL-SKG can be pursued in accordance with the estimations of the reciprocal channels between Alice and Bob. To be specific, we assign two full-row rank globally known pilot sequences, i.e., $\mathbf{P}_A\in\mathbb{C}^{N_A\times L_A}$ ($L_A\geq N_A$) for Alice, and $\mathbf{P}_B\in\mathbb{C}^{N_B\times L_B}$ for Bob ($L_B\geq N_B$). The received signals at Alice and Bob are:
\begin{equation}
\label{eq_4}
\begin{aligned}
    \mathbf{Y}_A&=\mathbf{H}_{BA}\cdot\mathbf{P}_B+\mathbf{N}_A,\\
    \mathbf{Y}_B&=\mathbf{H}_{AB}\cdot\mathbf{P}_A+\mathbf{N}_B,
\end{aligned}
\end{equation}
where $\mathbf{N}_A$ and $\mathbf{N_B}$ represent the noise matrices with i.i.d elements following complex Gaussian distribution, i.e., $\mathcal{CN}(0,\epsilon^2)$. Then, the reciprocal channels between them can be estimated by least squared (LS) method as:
\begin{equation}
\begin{aligned}
    vec(\hat{\mathbf{H}}_{AB})&=\left(\mathbf{P}_A^H\otimes\mathbf{I}_{N_B}\right)^\dag\cdot vec(\mathbf{Y}_B)\\
    vec(\hat{\mathbf{H}}_{BA})&=\left(\mathbf{P}_B^H\otimes\mathbf{I}_{N_A}\right)^\dag\cdot vec(\mathbf{Y}_A),
\end{aligned}
\end{equation}
where $vec(\cdot)$ is to vectorize a matrix, $\mathbf{I}_{N_A}$ ($\mathbf{I}_{N_B}$) is the identity matrix with size $N_A\times N_A$ ($N_B\times N_B$), and $(\cdot)^\dag$ is the Moore–Penrose inverse. After the channel estimation process, the secret key can be generated given the common elements of the reciprocal legitimate estimated channel $\hat{\mathbf{H}}_{AB}$ and $\hat{\mathbf{H}}_{BA}$. 

It is observed that the PL is enhanced by the randomly varied IRS phase, i.e., $\mathbf{w}$ in Eq. (\ref{eq1}). This makes the reciprocal legitimate channels $\mathbf{H}_{AB}$ and $\mathbf{H}_BA$ random for secret key generation, which thereby are hard to be interpreted by {\bf single/non-colluded} Eves.   

\subsection{Problem Formulation: A Worst Case of Colluded Eves}
In this part, we describe a worst case of colluded Eves, which are able to estimate the randomly changed IRS phase $\mathbf{w}$, and subsequently pave their way to construct the reciprocal channels $\mathbf{H}_{AB}$ and $\mathbf{H}_{BA}$ for decoding the secret key. We do so by firstly providing and explaining why single/non-colluded Eve is hard to decode PL secrete key, and then introduce the case of colluded Eves. Here, we assume a worst case where all direct channels $\mathbf{G}_{AR}$, $\mathbf{G}_{RB}$, $\mathbf{G}_{AB}$ are known to $M$ Eves. This is reasonable for the scarce randomness channels, given their (quasi) static properties over a long period. 

\subsubsection{Single/Non-Colluded Eve}
We first analyze the single Eve case. In the process Alice sends its pilot $\mathbf{P}_A$ to Bob, the received signal matrix at single Eve $m$ is:
\begin{equation}
\label{eve_received}
    \mathbf{Y}_{E_m}=\left(\mathbf{G}_{RE_m}\cdot diag(\mathbf{w})\cdot\mathbf{G}_{AR}+\mathbf{G}_{AE_m}\right)\cdot\mathbf{P}_A+\mathbf{N}_{E_m},
\end{equation}
where $\mathbf{N}_{E_m}$ represents the noise matrix with i.i.d elements following complex Gaussian distribution, i.e., $\mathcal{CN}(0,\epsilon^2)$. 
Given the known of direct static channel $\mathbf{G}_{AE_m}$ and pilot $\mathbf{P}_A$, we denote $\mathbf{Z}_{E_m}\triangleq\mathbf{Y}_{E_m}-\mathbf{G}_{AE_m}\mathbf{P}_A$. As such, Eq. (\ref{eve_received}) can be converted as:
\begin{equation}
\label{single_eve}
\begin{aligned}
    vec(\mathbf{Z}_{E_m})=&vec\left(\left(\mathbf{G}_{RE_m}\cdot diag(\mathbf{w})\cdot\mathbf{G}_{AR}\right)\cdot\mathbf{P}_A\right)+vec\left(\mathbf{N}_{E_m}\right)\\
    =&
    \begin{bmatrix}
    \mathbf{G}_{RE_m}\cdot diag(\mathbf{G_{AR}}\cdot\mathbf{p}_1)\\\vdots\\\mathbf{G}_{RE_m}\cdot diag(\mathbf{G_{AR}}\cdot\mathbf{p}_{L_A})
    \end{bmatrix}
    \cdot\mathbf{w}+vec(\mathbf{N}_{E_m}),
\end{aligned}
\end{equation}
where $\mathbf{P}_A=[\mathbf{p}_1,\cdots,\mathbf{p}_{L_A}]$.
It is noticed that Eq. (\ref{single_eve}) is an under-determined equation for $\mathbf{w}$, since
\begin{equation}
\label{single_eve1}
    rank\left(\begin{bmatrix}
    \mathbf{G}_{RE_m}\cdot diag(\mathbf{G_{AR}}\cdot\mathbf{p}_1)\\\vdots\\\mathbf{G}_{RE_m}\cdot diag(\mathbf{G_{AR}}\cdot\mathbf{p}_{L_A})
    \end{bmatrix}\right)<dim(\mathbf{w})=N_R.
\end{equation}
Eq. (\ref{single_eve1}) is because of the low-rank property of the narrow-band geometric channel $\mathbf{G}_{RE_m}$ \cite{9103231}. This indicates that the single Eve is difficult to estimate the IRS-induced randomness, and thereby hard to decode the PL-based secret key, given the lacking diversity (rank) of its received signals. Nevertheless, Eqs. (\ref{single_eve})-(\ref{single_eve1}) raise a straightforward idea to add colluded Eves, in order to increase the wiretap channel diversity/rank.

\subsubsection{Colluded Eves}
We then provide one case of colluded Eves, in which all $M$ Eves share their received signals. To simplify the expression, we denote:
\begin{equation}
    \bm{\Psi}_m\triangleq\begin{bmatrix}
    \mathbf{G}_{RE_m}\cdot diag(\mathbf{G_{AR}}\cdot\mathbf{p}_1)\\\vdots\\\mathbf{G}_{RE_m}\cdot diag(\mathbf{G_{AR}}\cdot\mathbf{p}_{L_A})
    \end{bmatrix}.
\end{equation}
Then, we establish an over-determined equation of the IRS-induced randomness $\mathbf{w}$ by stacking all $M$ Eves' received signals, i.e., 
\begin{equation}
\label{colluded}
    \begin{bmatrix}
    vec(\mathbf{Z}_{E_1})\\\vdots\\vec(\mathbf{Z}_{E_M})
    \end{bmatrix}
    =\begin{bmatrix}
    \bm{\Psi}_1\\\vdots\\\bm{\Psi}_M
    \end{bmatrix}
    \cdot\mathbf{w}+
    \begin{bmatrix}
    vec(\mathbf{N}_{E_1})\\\vdots\\vec(\mathbf{N}_{E_m})
    \end{bmatrix}.
\end{equation}
From Eq. (\ref{colluded}), one can estimate the IRS-induced randomness by the LS method, i.e., 
\begin{equation}
\label{w}
    \hat{\mathbf{w}}=\begin{bmatrix}
    \bm{\Psi}_1\\\vdots\\\bm{\Psi}_M
    \end{bmatrix}^\dag\cdot\begin{bmatrix}
    vec(\mathbf{Z}_{E_1})\\\vdots\\vec(\mathbf{Z}_{E_M})
    \end{bmatrix}. 
\end{equation}
As such, the reciprocal channels between Alice and Bob for PL-SKG can be estimated by taking $\hat{\mathbf{w}}$ into Eq. (\ref{eq1}), as we assume the known of other direct and static channels (i.e., $\mathbf{G}_{ab}$, $a\neq b\in\{A,R,B,E_1,\cdots,E_M\}$). From Eqs. (\ref{colluded})-(\ref{w}), we propose one worst case that weakens the PL based secret key. This is partially due to the global awareness of the pilot sequence for channel estimation, i.e., $\mathbf{P}_A$ and $\mathbf{P}_B$, which can also be used by Eves to estimate the legitimate channels and their related secret key. In the next section, we will introduce a new PL-SKG method, which does not require the globally known pilot sequence.

\section{Random Gaussian Matrix based PL-SKG}
In this section, we will elaborate our RGM PL-SKG method. Here, we avoid the usage of the globally known pilot sequence to extract the common properties of legitimate channels. Instead, Alice and Bob send random Gaussian matrices, and use the singular values of their received signal matrices to generate secret key. 

Before we start, we explain the compatibility of the designed RGM PL-SKG scheme with the wireless communication framework. The sketch of the designed RGM PL-SKG scheme is illustrated in Fig. \ref{figm1}(b). Here, we separate the PL-SKG from the wireless communication process. The channels for key generation are independent with that for further wireless communications. As such, in key generation part, one does not need to estimate the legitimate channels, since these channels will not be used in further wireless communications.

\subsection{Secrete Key Generation}
We assign the random Gaussian matrices sent by Alice and Bob as $\mathbf{X}_A\in\mathbb{C}^{N_A\times D}$ and $\mathbf{X}_B\in\mathbb{C}^{N_B\times D}$. Each element in $\mathbf{X}_A$ and in $\mathbf{X}_B$ is independent and follows the complex Gaussian distribution, i.e.,
\begin{equation}
\label{rgm11}
    \left(\mathbf{X}_A\right)_{m,n},~\left(\mathbf{X}_B\right)_{m,n}\sim\mathcal{CN}\left(0,\delta_{m,n}^2\right),~\sum_{n=1}^D\delta_{m,n}^2=C,
\end{equation}
where $C$ is the normalization constant. Then, we aim to show that the singular values of $\mathbf{H}_{AB}\mathbf{X}_A$ and of $\mathbf{H}_{BA}\mathbf{X}_B$, denoted as $\bm{\sigma}(\mathbf{H}_{AB}\mathbf{X}_A)$ and $\bm{\sigma}(\mathbf{H}_{BA}\mathbf{X}_B)$ have similar probability distribution functions (PDFs). We do so by following Theorems.  

\begin{Lem}
\label{theo1}
Consider a fixed matrix $\mathbf{H}\in\mathbb{C}^{N_B\times N_A}$, and a random matrix $\mathbf{X}\in\mathbb{C}^{N_A\times D}$, where each element is a random variable. Then, the singular values of $\mathbf{H}\mathbf{X}$, denoted as $\bm{\sigma}(\mathbf{H}\mathbf{X})$ has the same PDF with $\bm{\sigma}(\bm{\Xi}\mathbf{V}^H\mathbf{X})$, where $\mathbf{H}=\mathbf{U}\bm{\Xi}\mathbf{V}^H$ is the compact SVD. 
\end{Lem}
\begin{IEEEproof}
Consider a matrix $\mathbf{Q}$ and any orthogonal square matrix $\bm{\Gamma}$ (making $\bm{\Gamma}\mathbf{Q}$ feasible). Then, we have:
\begin{equation}
\label{l11}
    \bm{\sigma}(\bm{\Gamma}\cdot\mathbf{Q})=\bm{\sigma}(\mathbf{Q}).
\end{equation} 
Eq. (\ref{l11}) is because $(\bm{\Gamma}\mathbf{Q})^H\bm{\Gamma}\mathbf{Q}=\mathbf{Q}^H\bm{\Gamma}^H\bm{\Gamma}\mathbf{Q}=\mathbf{Q}^H\mathbf{Q}$. This suggests the identical eigenvalues of $(\bm{\Gamma}\mathbf{Q})^H\bm{\Gamma}\mathbf{Q}$ and $\mathbf{Q}^H\mathbf{Q}$, thereby the identical singular values of $\bm{\Gamma}\cdot\mathbf{Q}$ and $\mathbf{Q}$. 

We then perform the SVD on $\mathbf{H}$, i.e.,
\begin{equation}
\label{svdh}
    \mathbf{H}=\left[\mathbf{U}~\tilde{\mathbf{U}}\right]\cdot
    \begin{bmatrix}
    \bm{\Xi} & \mathbf{O}_{r\times(N_A-r)}\\
    \mathbf{O}_{(N_B-r)\times r} & \mathbf{O}_{(N_B-r)\times(N_A-r)}
    \end{bmatrix}
    \cdot
    \begin{bmatrix}
    \mathbf{V}^H\\\tilde{\mathbf{V}}^H.
    \end{bmatrix}
\end{equation}
Here, $\mathbf{U}$ of size $N_B\times r$ and $\tilde{\mathbf{U}}$ of size $N_B\times(N-r)$ compose the orthogonal left singular matrix. $\bm{\Xi}=diag([\xi_1,\cdots,\xi_r])$ with $ \xi_1\geq\cdots\geq\xi_r$ is the diagonal matrix of $r=rank(\mathbf{H})$ non-zero singular values. $\mathbf{V}$ of size $N_A\times r$ and ($\tilde{\mathbf{V}}$) of size $N_A\times(N_A-r)$ compose the orthogonal right singular matrix. As such, according to Eqs. (\ref{l11})-(\ref{svdh}), we have:
\begin{equation}
\label{proof1}
    \bm{\sigma}\left(\mathbf{H}\mathbf{X}\right)=\bm{\sigma}\left(\left[\mathbf{U}~\tilde{\mathbf{U}}\right]^H\cdot\mathbf{H}\mathbf{X}\right)=\bm{\sigma}\left(\bm{\Xi}\mathbf{V}^H\mathbf{X}\right)
\end{equation}
From Eq. (\ref{proof1}), we show that the singulars of $\mathbf{H}\mathbf{X}$ and $\bm{\Xi}\mathbf{V}^H\mathbf{X}$ are same, therefore following same probability distribution. 
\end{IEEEproof}

With the help of Lemma \ref{theo1}, we then analyze the PDF of $\bm{\sigma}(\mathbf{H}_{AB}\mathbf{X}_A)$.

\begin{Theo}
\label{theo2}
Consider a random Gaussian matrix $\mathbf{X}\in\mathbb{C}^{N_A\times D}$ following Eq. (\ref{rgm11}). Then, if $D$ is large (e.g., $D>50$), the largest singular value of $\mathbf{H}\mathbf{X}$, denoted as $\sigma_{max}(\mathbf{H}\mathbf{X})$, can be approximated as Gaussian distributed, i.e.,
\begin{equation}
\begin{aligned}
    &\sigma_{max}(\mathbf{H}\mathbf{X})\sim\mathcal{N}(\eta,~\iota^2)\\
    &\eta=\xi_1\cdot\left(4C^2-\sum_{n=1}^D\left(\sum_{m=1}^{N_A}\delta_{m,n}^2|V_{m,1}|^2\right)^2\right)^{\frac{1}{4}}\\
    &\iota^2=\xi_1^2\cdot\left(2C-\sqrt{4C^2-2\sum_{n=1}^D\left(\sum_{m=1}^{N_A}\delta_{m,n}^2|V_{m,1}|^2\right)^2}\right)
\end{aligned}
\end{equation}
where $\xi_1$ is the largest singular value of $\mathbf{H}$ and $V_{m,1}$ is the $(m,1)$th element of the right singular matrix $\mathbf{V}$ of $\mathbf{H}$.
\end{Theo}
\begin{IEEEproof}
We pursue SVD of $\mathbf{H}\mathbf{X}$, i.e.,
\begin{equation}
\label{append_svd}
    \sigma_{max}(\mathbf{H}\mathbf{X})=\bm{\varrho}_1^H\cdot\mathbf{H}\mathbf{X}\cdot\bm{\varpi}_1
\end{equation}
where $\bm{\varrho}_1$ ($\bm{\varpi}_1$) is the $1$th left (right) singular vector of $\mathbf{H}\mathbf{X}$, corresponding to the largest singular value $\sigma_{max}(\mathbf{H}\mathbf{X})$. Eq. (\ref{append_svd}) indicates that the singular value is a summation of random variables. Therefore, according to the central limit theorem (CLT), when the number of random variables (i.e., $D$) is large, we can deem $\sigma_{max}(\mathbf{H}\mathbf{X})$ follows the Gaussian distribution.

We next compute $\eta$ and $\iota$. From Lemma \ref{theo1}, given the compact SVD of $\mathbf{H}=\mathbf{U}\bm{\Xi}\mathbf{V}^H$, we have $\sigma_{max}(\mathbf{H}\mathbf{X})=\sigma_{max}(\bm{\Xi}\mathbf{V}^H\mathbf{X})$. This suggests:
\begin{equation}
    \sigma_{max}^2\left(\mathbf{H}\mathbf{X}\right)=\lambda_{max}\left(\bm{\Xi}\mathbf{V}^H\mathbf{X}\cdot\mathbf{X}^H\mathbf{V}\bm{\Xi}\right)
\end{equation}
where $\lambda_{max}(\cdot)$ represents the maximum eigenvalue of a matrix. Then, we notice that 
\begin{equation}
\label{append_mean}
\begin{aligned}
    &\mathbb{E}\left(\bm{\Xi}\mathbf{V}^H\mathbf{X}\cdot\mathbf{X}^H\mathbf{V}\bm{\Xi}\right)=\bm{\Xi}\mathbf{V}^H\cdot\mathbb{E}\left(\mathbf{X}\mathbf{X}^H\right)\cdot\mathbf{V}\bm{\Xi}\\
    =&\bm{\Xi}\mathbf{V}^H\cdot2\cdot diag\left(\left[\sum_{n=1}^D\delta_{1,n}^2,\cdots,\sum_{n=1}^D\delta_{N_A,n}^2\right]\right)\cdot\mathbf{V}\bm{\Xi}\\
    \overset{(\ref{rgm11})}{=}&2C\cdot\bm{\Xi}^2
\end{aligned}
\end{equation} 
where $\mathbb{E}(\cdot)$ denotes the expectation. Eq. (\ref{append_mean}) suggests that the expectation of $\bm{\Xi}\mathbf{V}^H\mathbf{X}\mathbf{X}^H\mathbf{V}\bm{\Xi}$ is a diagonal matrix. This thereby enables the approximation of the largest eigenvalue $\sigma_{max}^2(\mathbf{H}\mathbf{X})$ via the $(1,1)$th element in $\bm{\Xi}\mathbf{V}^H\mathbf{X}\mathbf{X}^H\mathbf{V}\bm{\Xi}$, i.e., 
\begin{equation}
\label{append_eig}
    \sigma_{max}^2\left(\mathbf{H}\mathbf{X}\right)\approx\xi_1^2\cdot\mathbf{V}_{:,1}^H\mathbf{X}\mathbf{X}^H\mathbf{V}_{:,1},
\end{equation}
where $\mathbf{V}_{:,1}$ is the first column of $\mathbf{V}$. 

According to Eq. (\ref{append_eig}), and the previous proof of approximated Gaussian distributed $\sigma_{max}(\mathbf{H}\mathbf{X})$, we can compute the mean and variance as:
\begin{align}
    &\eta^2+\iota^2=\mathbb{E}\left(\sigma_{max}^2\left(\mathbf{H}\mathbf{X}\right)\right)=2C\cdot\xi_1^2\label{eq37}\\
    &2\iota^4+4\eta^2\iota^2=\mathbb{D}\left(\sigma_{max}^2\left(\mathbf{H}\mathbf{X}\right)\right)\nonumber\\
    =&\xi_1^4\cdot \left(\mathbf{V}_{:,1}^T\otimes\mathbf{V}_{:,1}^H\right)\cdot Cov\left(vec(\mathbf{X}\mathbf{X}^H)\right)\cdot\left(\mathbf{V}_{:,1}^T\otimes\mathbf{V}_{:,1}^H\right)^H\nonumber\\
    =&4\xi_1^4\sum_{n=1}^D\left(\sum_{m=1}^{N_A}\delta_{m,n}^2V_{m,1}V_{m,1}^*\right)^2
    \label{eq38},
\end{align}
where $\mathbb{D}(\cdot)$ represents the variance, $V_{m,1}$ is the $(m,1)$th element of $\mathbf{V}$, and $(\cdot)^*$ is the conjugate operator. 
By solving the above two equations, $\eta$ and $\iota^2$ in this Theorem can be computed. 
\end{IEEEproof}

\begin{figure}[!t]
\centering
\includegraphics[width=3.5in]{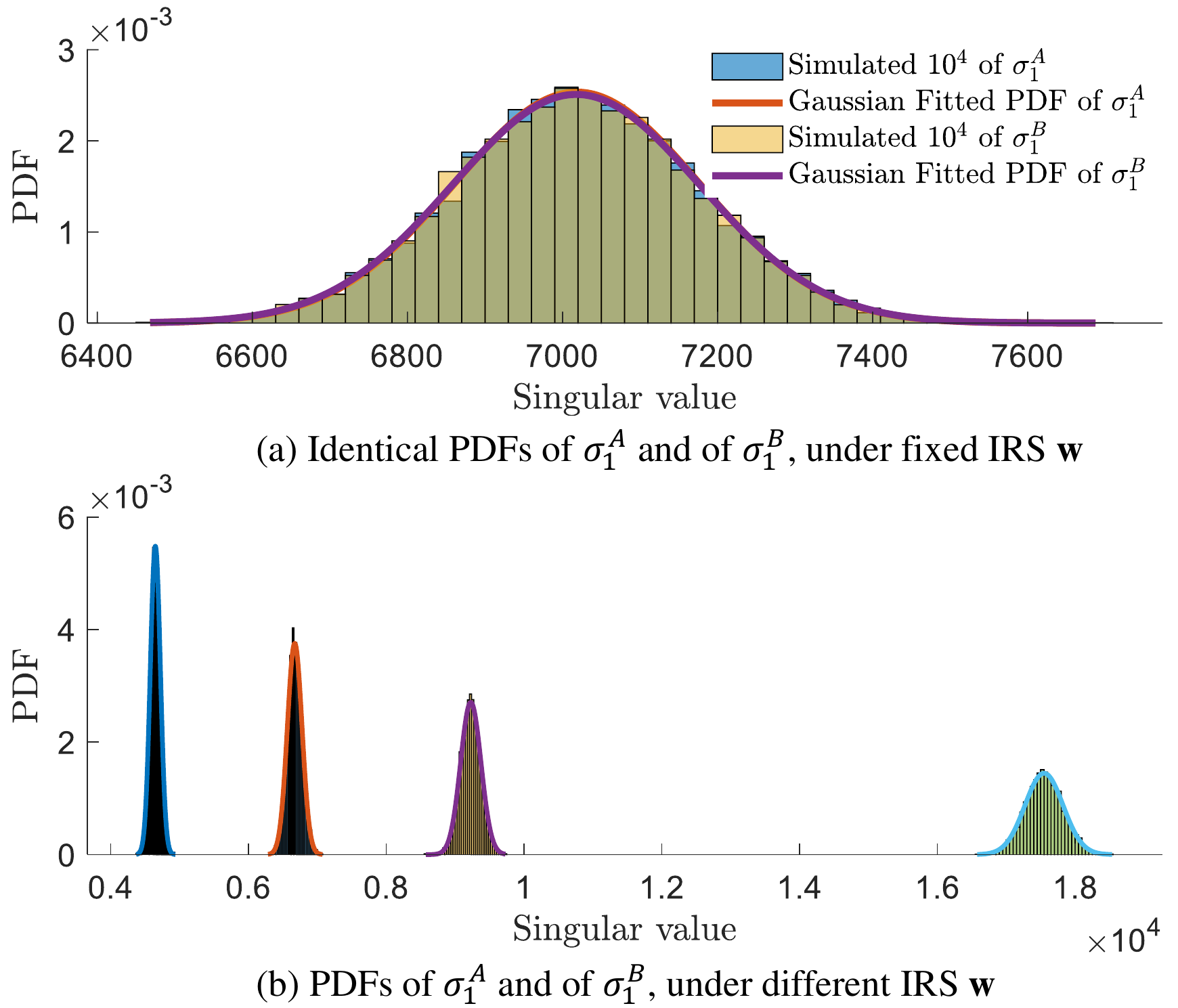}
\caption{Illustration of PDFs of $\sigma_1^A$ and of $\sigma_1^B$. (a) shows the similar PDFs of $\sigma_1^A$ and of $\sigma_1^B$ in one coherent time, i.e., with one fixed IRS $\mathbf{w}$. (b) shows the PDFs of $\sigma_1^A$ and of $\sigma_1^B$ under different coherent time, i.e., with different IRS $\mathbf{w}$. Combining (a) and (b), it is observed that $\sigma_1^A$ and of $\sigma_1^B$ are close under each coherent time, which demonstrates the feasibility of using $\sigma_1^A$ and of $\sigma_1^B$ for PL-SKG. }
\label{figm2}
\end{figure}

From Theorem \ref{theo2}, we can approximate the PDFs of $\sigma_{max}(\mathbf{H}_{AB}\mathbf{X}_A)$ and of $\sigma_{max}(\mathbf{H}_{BA}\mathbf{X}_B)$. One illustration of such PDFs are provided in Fig. \ref{figm2}(a), which demonstrates our Theorem \ref{theo2}, and also shows similar shapes of the two PDFs. We then describe such similar PDFs via the following Proposition. 

\begin{Prop}
\label{prop1}
Denote $\sigma_{max}(\mathbf{H}_{AB}\mathbf{X}_A)\sim\mathcal{N}(\eta_B,\iota_B^2)$ and $\sigma_{max}(\mathbf{H}_{BA}\mathbf{X}_B)\sim\mathcal{N}(\eta_A,\iota_A^2)$. Then, 
\begin{equation}
\begin{aligned}
&~\eta_{min}\leq\eta_A,\eta_B\leq\eta_{max}\\
    \eta_{min}&=\xi_1\left(4C^2-\sum_{n=1}^D\max_{m}\delta_{m,n}^4\right)^\frac{1}{4}\\
    \eta_{max}&=\xi_1\left(4C^2-\sum_{n=1}^D\min_{m}\delta_{m,n}^4\right)^\frac{1}{4},
\end{aligned}
\end{equation}
\begin{equation}
\begin{aligned}
    &~~~~~~~~~~\iota^2_{min}\leq\iota_A^2,\iota_B^2\leq\iota^2_{max}\\
    \iota_{min}^2&=\xi_1^2\cdot\left(2C-\sqrt{4C^2-2\sum_{n=1}^D\min_{m}\delta_{m,n}^4}\right)\\
    \iota_{max}^2&=\xi_1^2\cdot\left(2C-\sqrt{4C^2-2\sum_{n=1}^D\max_{m}\delta_{m,n}^4}\right).
\end{aligned}
\end{equation}
\end{Prop}
\begin{IEEEproof}
The proof can be easily pursued by noticing $\sum_{n=1}^D\min_{m}\delta_{m,n}^4\leq\sum_{n=1}^D(\sum_{m=1}^{N_A}\delta_{m,n}^2|V_{m,1}|^2)^2\leq\sum_{n=1}^D\max_{m}\delta_{m,n}^4$.
\end{IEEEproof}

With the help of Proposition \ref{prop1} and Fig. \ref{figm2}, $\sigma_{max}(\mathbf{H}_{AB}\mathbf{X}_A)$ and $\sigma_{max}(\mathbf{H}_{BA}\mathbf{X}_B)$ can be demonstrated to have similar PDFs.
Leveraging this, we hereby design the random Gaussian matrix based secrete key, using the received signal's singular values of Alice and Bob. The RGM PL-SKG process are pursued by the repetition of following 3 steps.

{\bf Step 1:} To induce the random channel fading for key generation, the phase of IRS phase controller $\mathbf{w}=[\exp(j\theta_1),\cdots,\exp(j\theta_{N_R})]^T$ is evenly selected from $[0,2\pi]^{N_R}$. This $\mathbf{w}$ will be invariant during the coherent time, or saying in each iteration of Steps. 1-3. 

{\bf Step 2:}  Alice and Bob generate random  Gaussian matrices according to Eq. (\ref{rgm11}), and send to each other.
Here, except for the transmitter, all nodes (e.g., IRS and Eves) do not know the exact form of $\mathbf{X}_A$ and $\mathbf{X}_B$, which prevents the estimation of legitimate channels from Eves. 

{\bf Step 3: } Alice and Bob receive the signal matrices as $\mathbf{H}_{BA}\cdot\mathbf{X}_B+\mathbf{N}_A$ and $\mathbf{H}_{AB}\cdot\mathbf{X}_A+\mathbf{N}_B$ respectively. They derive the largest singular values as $\sigma_{max}(\mathbf{H}_{BA}\mathbf{X}_B+\mathbf{N}_A)$ and $\sigma_{max}(\mathbf{H}_{AB}\mathbf{X}_A+\mathbf{N}_B)$. Leveraging the common singular values, Alice and Bob can use quantization techniques \cite{5422766} to map them into the binary secret key.

After the description of the proposed RGM PL-SKG, we highlight the advantages in the following, which have 2 folds. First, the scheme can to some extent address the scarce randomness induced secret key shortage challenge in PL SKG. This is attributed the combination of (i) the separation of SKG from communication, and (ii) the IRS-induced randomness. For example, before communication process, IRS can create enough randomness and fading channels for our proposed SKG, without interfering the following communication channel statistics. 
Second, the avoidance of the globally known pilot sequence is able to prevent the illegal channel probing and subsequently the secret key decoding by the Eves. Here, we express the received signal by colluded Eves, i.e.,
\begin{equation}
\label{eve_}
\begin{aligned}
    &~~~~~~~~~~~~~~~~~~~~~~~\mathbf{Z}_E=\mathbf{H}_{eve}\cdot\mathbf{X}_A+\mathbf{N}_E\\
    &\mathbf{H}_{eve}\triangleq
    \begin{bmatrix}
    \mathbf{G}_{RE_1}\cdot diag(\mathbf{w})\cdot\mathbf{G}_{AR}+\mathbf{G}_{AE_1}\\\vdots\\\mathbf{G}_{RE_M}\cdot diag(\mathbf{w})\cdot\mathbf{G}_{AR}+\mathbf{G}_{AE_M}
    \end{bmatrix},
    \mathbf{N}_E\triangleq
    \begin{bmatrix}
    \mathbf{N}_{E_1}\\\vdots\\\mathbf{N}_{E_M}
    \end{bmatrix}.
\end{aligned}
\end{equation}
It is seen from Eq. (\ref{eve_}) that the unknown random Gaussian matrix $\mathbf{X}_A$ prevent the estimation of neither IRS's random phase $\mathbf{w}$ nor the common singular values.
We will further analyze the rate of secret key in the following parts.

\subsection{Secret Key Rate Analysis}
In this part, we analyze the SKR of using the largest common singular value. To simplify the denotation, we denote $\sigma_1^A$ ($\sigma_1^B$) as the largest singular value of Alice's (Bob's) received signals, i.e.,
\begin{equation}
\label{deff}
\begin{aligned}
    \sigma_1^A&\triangleq\sigma_{max}\left(\mathbf{H}_{BA}\cdot\mathbf{X}_B+\mathbf{N}_A\right)\\
    \sigma_1^B&\triangleq\sigma_{max}\left(\mathbf{H}_{AB}\cdot\mathbf{X}_A+\mathbf{N}_B\right),
\end{aligned}
\end{equation} 

The SKR can be expressed via its lower-bound, i.e., \cite{7299681,243431,256484,5756230}
\begin{equation}
\label{rate}
    R_s=I\left(\sigma_1^A;\sigma_1^B\right)-\max\Big\{I\left(\sigma_1^A;\mathbf{Z}_E\right),
    I\left(\sigma_1^B;\mathbf{Z}_E\right)\Big\}, 
\end{equation}
where $I(\sigma_1^A;\sigma_1^B)$ is the mutual information of common singular values, and $I(\sigma_1^A;\mathbf{Z}_E)$ or $I(\sigma_1^B;\mathbf{Z}_E)$ is the secret key leakage. We will compute them separately in the following. 

\subsubsection{Mutual Information of Common Singular Values}
We firstly compute $I(\sigma_1^A;\sigma_1^B)$ by:
\begin{equation}
\label{rate1}
    I\left(\sigma_1^A;\sigma_1^B\right)=\iint p\left(\sigma_1^A,\sigma_1^B\right)\log_2\frac{p\left(\sigma_1^A,\sigma_1^B\right)}{p\left(\sigma_1^A\right),p\left(\sigma_1^B\right)}d\sigma_1^Ad\sigma_1^B.
\end{equation}
where $p(\cdot)$ represents the PDF.

The PDFs of $\sigma_1^A$ and $\sigma_1^B$, and their joint distribution can be expressed as:
\begin{equation}
\label{pdf1}
p\left(\sigma_1^A\right)=\sum_{\mathbf{w}\in\mathcal{K}^{N_R}} p(\mathbf{w})\cdot p\left(\sigma_1^A|\mathbf{w}\right),
\end{equation}
\begin{equation}
\label{pdf2}
p\left(\sigma_1^B\right)=\sum_{\mathbf{w}\in\mathcal{K}^{N_R}}p(\mathbf{w})\cdot p\left(\sigma_1^B|\mathbf{w}\right),
\end{equation}
\begin{equation}
\label{pdf3}
\begin{aligned}
    p\left(\sigma_1^A,\sigma_1^B\right)&=\sum_{\mathbf{w}\in\mathcal{K}^{N_R}} p(\mathbf{w})\cdot p\left(\sigma_1^A,\sigma_1^B|\mathbf{w}\right)\\ &=\sum_{\mathbf{w}\in\mathcal{K}^{N_R}} p(\mathbf{w})\cdot p\left(\sigma_1^A|\mathbf{w}\right)\cdot p\left(\sigma_1^B|\mathbf{w}\right).
\end{aligned}
\end{equation}
In Eqs. (\ref{pdf1})-(\ref{pdf3}), $p(\mathbf{w})$ is the probability distribution of the IRS phase shift $\mathbf{w}$. As we independently and evenly select each phase of $\mathbf{w}$ in $[0,2\pi]$, $p(\mathbf{w})=1/(2\pi)^{N_R}$. $p(\sigma_1^A|\mathbf{w})$ and $p(\sigma_1^B|\mathbf{w})$ are the PDFs of singular values conditioned on the fixed $\mathbf{w}$. The computations of exact form of $p(\sigma_1^A|\mathbf{w})$ and $p(\sigma_1^B|\mathbf{w})$ are difficult, due to the Gaussian distribution of each element with non-zero means \cite{shen2001singular,edelman200518}. We hereby approximate these conditional PDFs via following Proposition.

\begin{Prop}
\label{prop2}
Given one fixed $\mathbf{w}$, the PDFs of $\sigma_1^B$ and $\sigma_1^A$ in Eq. (\ref{deff}) can be approximated as Gaussian type, i.e.,
\begin{equation}
\label{approx_pdf}
\begin{aligned}
    p\left(\sigma_1^B|\mathbf{w}\right)&\approx\mathcal{N}\left(\sigma_1^B;~\mu_B,~\varsigma_B^2\right),\\
    \mu_B=&\Bigg(4\left(C\xi_1^2+D\epsilon^2\right)^2-4C\xi_1^2\epsilon^2-2D\epsilon^4\\
    &-2\xi_1^4\sum_{n=1}^D\left(\sum_{m=1}^{N_A}\delta_{m,n}^2|V_{m,1}|^2\right)^2\Bigg)^\frac{1}{4},\\
    \varsigma_B^2=&2C\xi_1^2+2D\epsilon^2-\mu_B^2
\end{aligned}
\end{equation}
\begin{equation}
\label{approx_pdf1}
\begin{aligned}
    p\left(\sigma_1^A|\mathbf{w}\right)&\approx\mathcal{N}\left(\sigma_1^A;~\mu_A,~\varsigma_A^2\right),\\
    \mu_A=&\Bigg(4\left(C\xi_1^2+D\epsilon^2\right)^2-4C\xi_1^2\epsilon^2-2D\epsilon^4\\
    &-2\xi_1^4\sum_{n=1}^D\left(\sum_{m=1}^{N_B}\delta_{m,n}^2|U_{m,1}|^2\right)^2\Bigg)^\frac{1}{4},\\
    \varsigma_A^2=&2C\xi_1^2+2D\epsilon^2-\mu_A^2
\end{aligned}
\end{equation}
where $\xi_1=\sigma_{max}(\mathbf{H}_{AB})=\sigma_{max}(\mathbf{H}_{BA})$, $\epsilon^2$ is the elemental-wise variance of additive noise matrices (i.e., $\mathbf{N}_B$ and $\mathbf{N}_A$) , and $U_{m,1}$ ($V_{m,1}$) is the $(m,1)$th element of the left (right) singular matrix of $\mathbf{H}_{AB}$.
\end{Prop}
\begin{IEEEproof}
Here, we explain the PDF computation of $\sigma_1^B|\mathbf{w}$. This is similar to $\sigma_1^A|\mathbf{w}$. We firstly show that $\sigma_1^B|\mathbf{w}$ can be approximated as Gaussian distributed. This is due to that, given fixed $\mathbf{w}$, $\sigma_1^B$ can be expressed as the summation of random variables, i.e.,
\begin{equation}
    \sigma_1^B=\bm{\vartheta}_1^H\cdot\left(\mathbf{H}_{AB}\mathbf{X}_A+\mathbf{N}_B\right)\cdot\bm{\kappa}_1
\end{equation}
where $\bm{\vartheta}_1$ ($\bm{\kappa}_1$) is the left (right) singular vector of $\mathbf{H}_{AB}\mathbf{X}_A+\mathbf{N}_B$, corresponding to the largest singular value. As such, according to the CLT, $\sigma_1^B|\mathbf{w}$ can be approximated by Gaussian distribution (when $D$ is large), and we can write:
\begin{equation}
\label{append_pdf2}
    \sigma_1^B|\mathbf{w}\sim\mathcal{N}\left(\mu_B,~\varsigma_B^2\right).
\end{equation}

Then, from Lemma \ref{theo1}, we have
\begin{equation}
\begin{aligned}
    \sigma_1^B=\sigma_1\left(\bm{\Delta}+\mathbf{N}\right),~
    \bm{\Delta}=\begin{bmatrix}
    \bm{\Pi} & \mathbf{O}_{r\times(D-r)}\\
    \mathbf{O}_{(N_B-r)\times r} & \mathbf{O}_{(N_B-r)\times(D-r)}
    \end{bmatrix}
\end{aligned}
\end{equation}
where $\bm{\Pi}$ is diagonal matrix composed by $r$ non-zero singular values of $\mathbf{H}_{AB}\mathbf{X}_A$, and $\mathbf{N}\sim\mathcal{CN}(0,\epsilon^2)^{N_B\times N_B}$ is the transformed noise matrix by orthogonal square matrices. As such, 
\begin{equation}
    \left(\sigma_1^B\right)^2=\lambda_{max}\left(\bm{\Delta}\bm{\Delta}+\mathbf{N}\bm{\Delta}+\bm{\Delta}\mathbf{N}^H+\mathbf{N}\mathbf{N}^H\right).
\end{equation}
It is noticed that the expectation of $\bm{\Delta}\bm{\Delta}+\mathbf{N}\bm{\Delta}+\bm{\Delta}\mathbf{N}^H+\mathbf{N}\mathbf{N}^H$ is a diagonal matrix. This therefore enables the approximation of $(\sigma_1^B)^2$ by the $(1,1)$th element, i.e., 
\begin{equation}
\label{append_approx2}
\begin{aligned}
    \left(\sigma_1^B\right)^2\approx&\zeta^2+\zeta\cdot\left(N_{1,1}+N_{1,1}^*\right)+\sum_{n=1}^DN_{1,n}N_{1,n}^*\\
    =&\zeta^2+2\zeta\cdot Re[N_{1,1}]+\sum_{n=1}^DRe^2[N_{1,n}]+Im^2[N_{1,n}],
\end{aligned}
\end{equation}
where $\zeta\triangleq\sigma_{max}(\mathbf{H}_{BA}\mathbf{X}_B)$, and $N_{1,n}$ is the $(1,n)$th element of $\mathbf{N}$. 

Taking Eq. (\ref{eq37}) from Theorem \ref{theo2} into Eq. (\ref{append_approx2}), the expectation and variance of $(\sigma_1^B)^2$ under fixed $\mathbf{w}$ can be computed:
\begin{equation}
\label{eq46}
\begin{aligned}
    \mu^2+\varsigma^2=&\mathbb{E}\left(\left(\sigma_1^B|\mathbf{w}\right)^2\right)=\mathbb{E}\left(\zeta^2|\mathbf{w}\right)+2\mathbb{E}(\zeta|\mathbf{w})\mathbb{E}\left(Re[N_{1,1}]\right)\\
    &+\sum_{i=1}^D\left(\mathbb{E}\left(Re^2[N_{1,i}]\right)+\mathbb{E}\left(Im^2[N_{1,i}]\right)\right)\\
    =&2C\xi_1^2+2D\epsilon^2,
\end{aligned}
\end{equation}
\begin{equation}
\label{eq47}
\begin{aligned}
    2\varsigma^4+4\mu^2\varsigma^2=&\mathbb{D}\left(\left(\sigma_1^B|\mathbf{w}\right)^2\right)=\mathbb{D}\left(\zeta^2|\mathbf{w}\right)+4\mathbb{D}\left(\zeta\cdot Re[N_{1,1}]|\mathbf{w}\right)\\
    &+\sum_{n=1}^D\left(\mathbb{D}\left(Re^2[N_{1,n}]\right)+\mathbb{D}\left(Im^2[N_{1,n}]\right)\right)\\
    =&4\xi_1^4\sum_{n=1}^D\left(\sum_{i=1}^{N_A}\delta^2_{i,n}|V_{i,1}|^2\right)^2+8C\xi_1^2\epsilon^2+4D\epsilon^4,
\end{aligned}
\end{equation}
Then, by solving Eqs. (\ref{eq46})-(\ref{eq47}), we derive the mean and variance, which are shown in Eq. (\ref{approx_pdf}). 
\end{IEEEproof}

With the help of Proposition \ref{prop2}, the mutual information of common singular value can be derived, by numerically computing Eq. (\ref{pdf1}), Eq. (\ref{pdf2}) and Eq. (\ref{pdf3}), and taking them into Eq. (\ref{rate1}).

\subsubsection{Secret Key Leakage Rate}
We then analyze the secret key leakage rate, i.e., $I(\sigma_1^A;\mathbf{Z}_E)$ or $I(\sigma_1^B;\mathbf{Z}_E)$ in Eq. (\ref{rate}). Here, we consider an upper-bound of $I(\sigma_1^A;\mathbf{Z}_E)$ which is the same of $I(\sigma_1^B;\mathbf{Z}_E)$, i.e.,
\begin{equation}
\label{leakage}
\begin{aligned}
    &I\left(\sigma_1^A;\mathbf{Z}_E\right)=I\left(\mathbf{Z}_E;\sigma_1^A;\right)\\
    =&I\left(\mathbf{Z}_E;\sigma_1^A,\mathbf{w}\right)-I\left(\mathbf{Z}_E;\mathbf{w}|\sigma_1^A\right)\\
    =&I\left(\mathbf{Z}_E;\mathbf{w}\right)+I\left(\mathbf{Z}_E;\sigma_1^A|\mathbf{w}\right)-I\left(\mathbf{Z}_E;\mathbf{w}|\sigma_1^A\right)\\
    =&I\left(\mathbf{Z}_E;\mathbf{w}\right)+H\left(\mathbf{Z}_E|\mathbf{w}\right)-H\left(\mathbf{Z}_E|\sigma_1^A\right)\\
    \overset{(a)}{<}&I\left(\mathbf{Z}_E;\mathbf{w}\right).
\end{aligned}
\end{equation}
where $H(\cdot)$ represents the information entropy. In Eq. (\ref{leakage}), (a) holds for the fact that $\mathbf{Z}_E$ remains less uncertainty under the known of $\mathbf{w}$ compared to $\sigma_1^A$, given Eq. (\ref{eve_}). 
From Eq. (\ref{leakage}), we can estimate the secret key leakage rate via its upper-bound, which can be expressed as:
\begin{equation}
\label{up_leakage}
\begin{aligned}
    I\left(\mathbf{Z}_E;\mathbf{w}\right)=&\sum_{w\in\mathcal{K}^{N_R}}\int_{\substack{Re[\mathbf{Z}_E],\\Im[\mathbf{Z}_E]\\\in\mathbb{R}^{M\times N_E}}} p\left(\mathbf{Z}_E,\mathbf{w}\right)\\
    &\cdot\log_2\frac{p\left(\mathbf{Z}_E,\mathbf{w}\right)}{p\left(\mathbf{Z}_E\right)P\left(\mathbf{w}\right)}dRe[\mathbf{Z}_E]dIm[\mathbf{Z}_E].
\end{aligned}
\end{equation}
In Eq. (\ref{up_leakage}), $p(\mathbf{Z}_E,\mathbf{w})$ is the joint probability distribution of $\mathbf{Z}_E$ and $\mathbf{w}$. According to Eq. (\ref{eve_}), we have:
\begin{equation}
\label{up_leakage1}
    p\left(\mathbf{Z}_E,\mathbf{w}\right)=\int_{\substack{Re[\mathbf{X}_A],\\Im[\mathbf{X}_A]\\\in\mathbb{R}^{N_A\times D}}}p\left(\mathbf{X}_A\right)p\left(\mathbf{Z}_E|\mathbf{X}_A,\mathbf{w}\right)dRe[\mathbf{X}_A]dIm[\mathbf{X}_A],
\end{equation}
\begin{equation}
\label{up_leakage2}
    p\left(\mathbf{Z}_E\right)=\sum_{\mathbf{w}\in\mathcal{K}^{N_R}}p(\mathbf{w})\cdot p\left(\mathbf{Z}_E,\mathbf{w}\right),
\end{equation}
where $p(\mathbf{X}_A)$ is computed according to Eq. (\ref{rgm11}). $p\left(\mathbf{Z}_E|\mathbf{X}_A,\mathbf{w}\right)$ can be expressed according to Eq. (\ref{eve_}), i.e, 
\begin{equation}
\label{up_leakage3}
\begin{aligned}
    p&\left(\mathbf{Z}_E|\mathbf{X}_A,\mathbf{w}\right)=\frac{1}{\sqrt{(2\pi)^{N_EM\times D}}}\\
    &\cdot\exp\left(-\frac{tr\left(\left(\mathbf{Z}_E-\mathbf{H}_{eve}\mathbf{X}_A\right)^H\left(\mathbf{Z}_E-\mathbf{H}_{eve}\mathbf{X}_A\right)\right)}{2\epsilon^2}\right).
\end{aligned}
\end{equation}

As such, by taking Eqs. (\ref{up_leakage1})-(\ref{up_leakage3}) into Eq. (\ref{up_leakage}), we derive the upper-bound of secret key leakage rate, which combined with Eq. (\ref{rate1}), can be used to compute the low-bound of SKR.

\section{Numerical Simulations}
In this section, we evaluate our proposed RGM PL-SKG via the simulated SKR. The configuration of the IRS assisted scarce randomness mmWave communication system is provided as follows. The numbers of antennas for Alice, Bob and each Eve are assigned as $N_A=N_B=N_E=16$ (all considered as ULA). The number of IRS elements (UPA) is $N_R=N_{R,x}\times N_{R,y}=10\times 10=100$. We range the number of colluded Eves, i.e., $M$, from $1$ to $100$ to see the comparison of the proposed RGM PL-SKG and the global known pilot based methods. For all direct channels $\mathbf{G}_{ab}$ ($a\neq b\in\{A,R,B,E_1,\cdots,E_M\}$), we use the narrow-band geometric channel model in Eq. (\ref{direct_channels}). Here, the number of paths $L_{ab}$ are set as random integer between $1$ to $10$. The azimuth (elevation) AoA and AoD (i.e., $\alpha_{ab,l}$, $\beta_{ab,l}$, $\theta_{ab,l}$, and $\gamma_{ab,l}$) are randomly selected from $[0,2\pi]$. The complex gain $g_{ab,l}$ are set as $g_{ab,1}=1$ for line-of-sight (LoS) path, and as complex Gaussian variables for other non-line-of-sight (NLoS) paths. We use the information theoretical estimators (ITE) Toolbox \cite{szabo2014information} to simulate the mutual information and secret key rate. 

For comparison, we select the RSS based PL-SKG in \cite{9442833,staat2020intelligent}, using the globally known pilot and LS for channel estimation. Here, we abbreviate it as the pilot based PL-SKG. It is noteworthy that other pilot based SKG (e.g., with ZF and MMSE channel estimators) will have similar results, since they cannot prevent the colluded Eves proposed in Section II. C to estimate the legitimate channel. 

\subsection{Performance of Colluded Eves}

\begin{figure}[!t]
\centering
\includegraphics[width=3.5in]{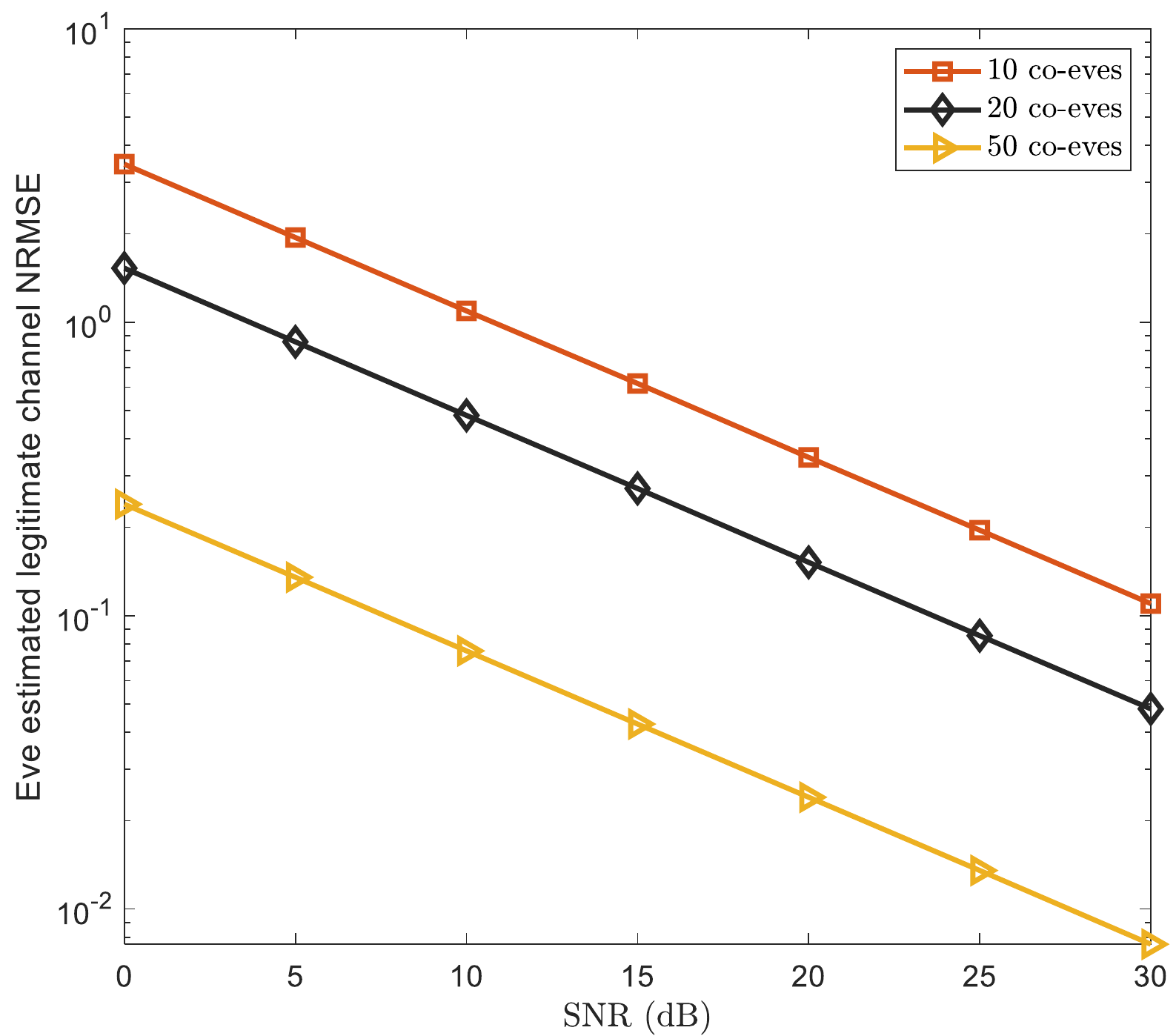}
\caption{Colluded Eves' estimation of IRS-assisted legitimate channels, via the globally known pilot sequence. It is seen that as the number of Eves increases, the NRMSE of legitimate channel estimation decreases. This shows the ability of our provided colluded Eve scheme for illegal channel probing via the globally known pilot.}
\label{eve_estimate}
\end{figure}

We firstly evaluate the estimation accuracy of the legitimate channels by the colluded Eves proposed in Section II. C.  Here, we use the normalised root mean squared error (NRMSE) to describe the estimation accuracy, i.e.,
\begin{equation}
    NRMSE=\frac{\|\hat{\mathbf{H}}_{AB}-\mathbf{H}_{AB}\|_{fro}}{\|\mathbf{H}_{AB}\|_{fro}}
\end{equation}
where $\hat{\mathbf{H}}_{AB}$ is the Eves' estimation of legitimate channel using Eq. (\ref{w}), and $\|\cdot\|_{fro}$ denotes the Frobenius norm. 

The results are shown in Fig. \ref{eve_estimate}. It is observed that the NRMSE is decreasing with the increasing of either the number of colluded Eves (i.e., $M$), or the SNR. When $M=50$ Eves are used, the NRMSE approaches to $1\%$ in large SNR regime (e.g., SNR$>25$dB), which suggests an accurate channel estimation for further illegally decoding the secret key.   
This attack is due to the usage of the globally known pilot sequences for legitimate users' channel estimation, which however paves the way for the colluded Eves to estimate the IRS random phase $\mathbf{w}$ and subsequently the legitimate channel $\mathbf{H}_{AB}$.

\subsection{Performance of Proposed RGM PL-SKG}
\begin{figure}[!t]
\centering
\includegraphics[width=3.5in]{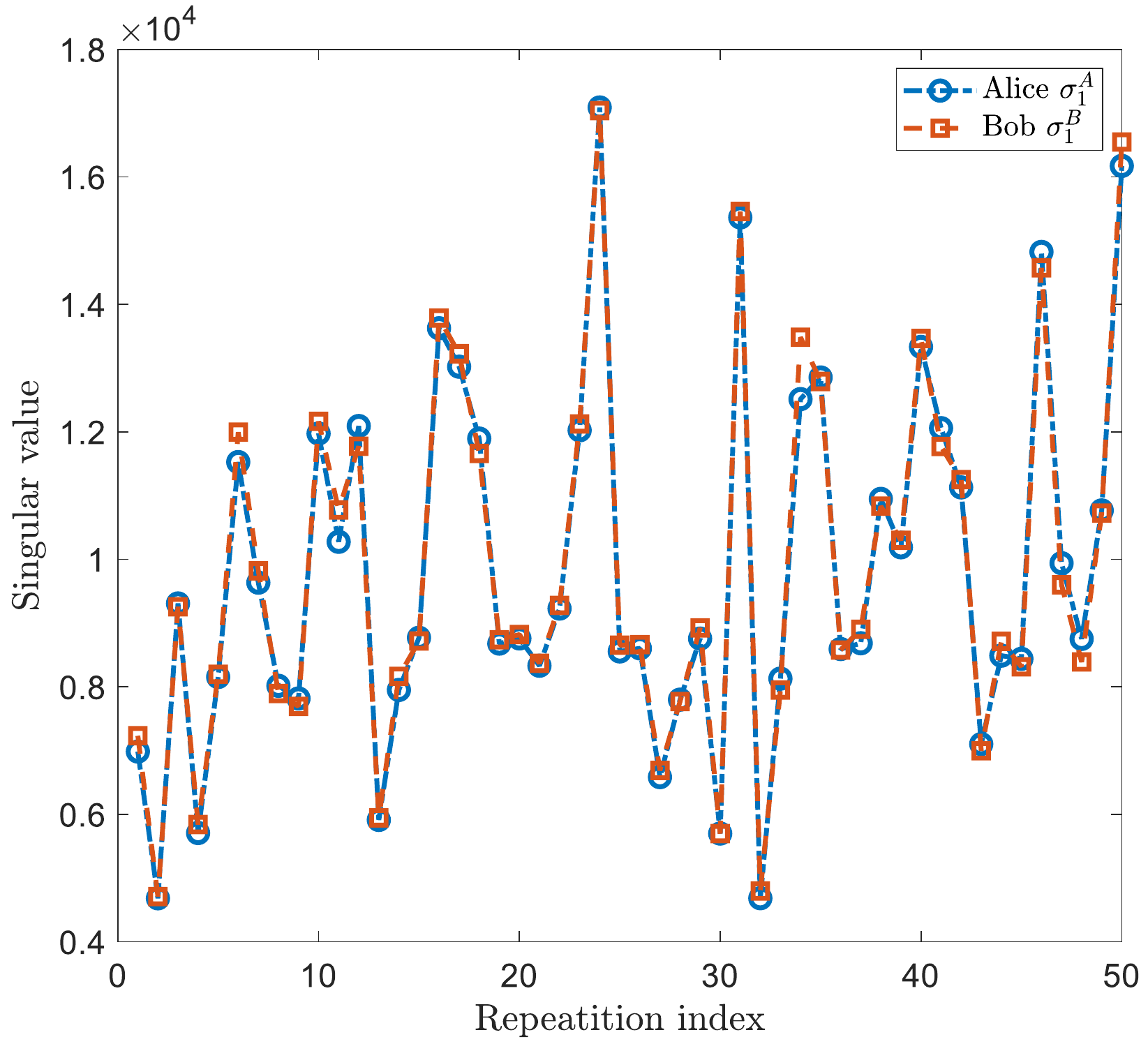}
\caption{Illustration of singular values of Alice's and Bob's received matrices. Their common singular values suggest the potentiality for secret key generation. }
\label{common_sv}
\end{figure}

We then evaluate the performance of our proposed RGM PL-SKG. We firstly show one illustration of the common singular values of Alice's and Bob's received matrices, when sending the designed Gaussian matrix in Eq. (\ref{rgm11}) instead of the globally known pilot sequence. In Fig. \ref{common_sv}, the x-coordinate represents the index of repetition of the three steps in RGM PL-SKG, and the y-coordinate is the singular value. In each repetition (coherent duration), the IRS generate one random phase $\mathbf{w}$, which will be unchanged until the next repetition. It is seen from Fig. \ref{common_sv} that the singular values of Alice and Bob (i.e., $\sigma_1^A$, and $\sigma_1^B$) are always close to each other, which thereby enables the further quantization process for secret key generation. 

\begin{figure}[!t]
\centering
\includegraphics[width=3.5in]{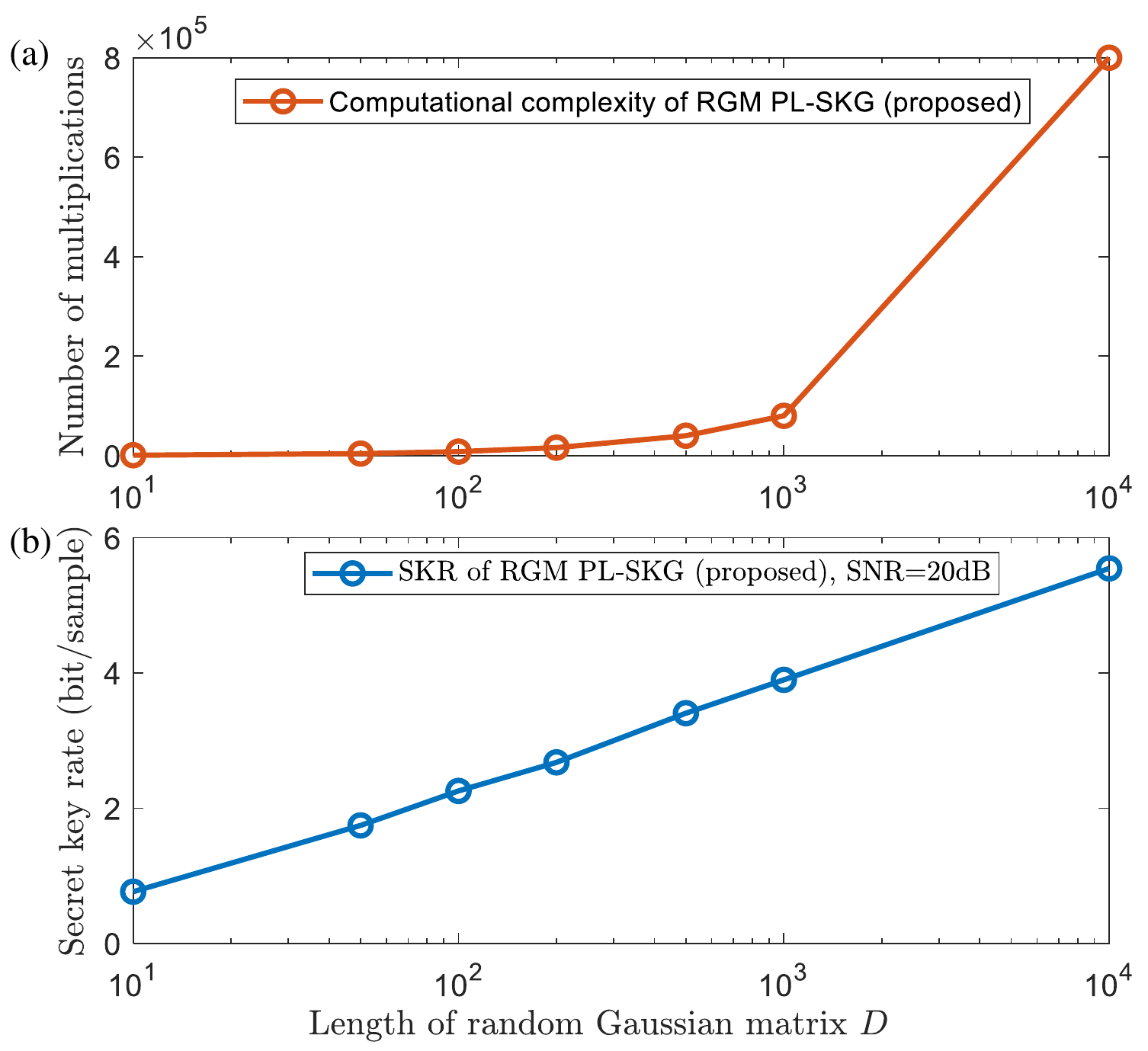}
\caption{SKR and computational complexity of our proposed RGM PL-SKG versus the length of random Gaussian matrix, i.e., $D$.}
\label{skr_lenrandom}
\end{figure}

We next provide the SKR and the computational complexity of our proposed RGM PL-SKG, versus the number of columns of the sent Gaussian matrix in Fig. \ref{skr_lenrandom}. Here, the computational complexity is measured as the number of multiplications. In our proposed RGM PL-SKG, this number equals that of SVD which is $O(N_A^2\cdot D)$ for Alice and $O(N_B^2\cdot D)$ for Bob. As such, in Fig. \ref{skr_lenrandom} (a), the number of multiplications in our proposed scheme is linear and increases with the growth of the length of the sent Gaussian matrix $D$. Then, Fig. \ref{skr_lenrandom}(b) shows an increased SKR of the proposed scheme with the growth of the length of the sent random Gaussian matrix $D$. This is because a larger random matrix sent by Alice and Bob can help reduce the information of the IRS phase $\mathbf{w}$ from Eves' received signals, i.e., lowering the mutual information $I(\mathbf{Z}_E; \mathbf{w})$.

\subsection{SKR Comparisons}

\begin{figure}[!t]
\centering
\includegraphics[width=3.5in]{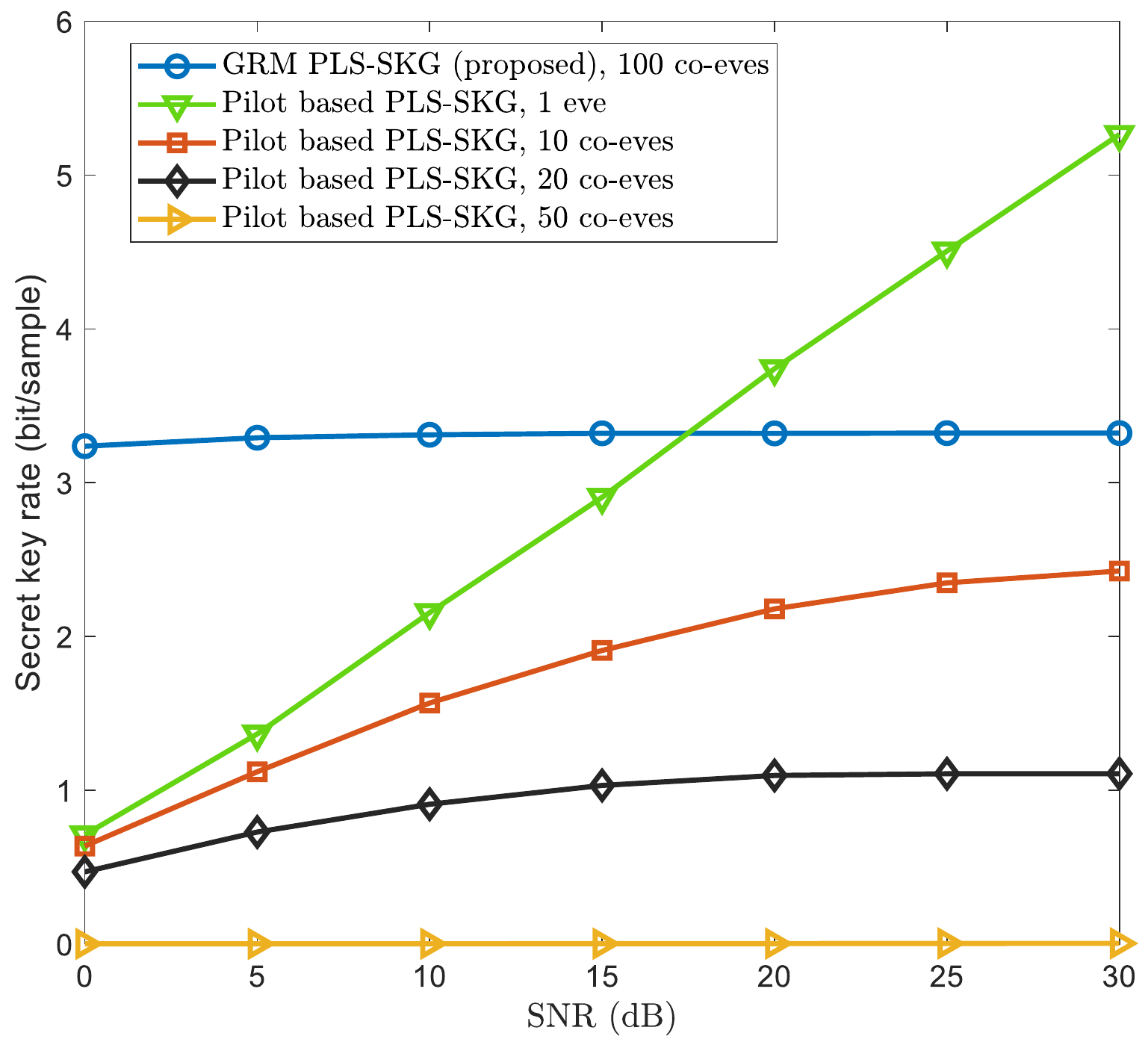}
\caption{The comparisons of SKRs against colluded Eves. It is shown that our proposed RGM PL-SKG (i) has an outstanding SKR in the low SNR region, and (ii) out-performing SKR over the current pilot based PL-SKG.}
\label{skr_compare}
\end{figure}

We finally compare the SKRs of our proposed RGM PL-SKG with the globally known pilot based PL-SKG method. We describe and explain the results in Fig. \ref{skr_compare} from three aspects. 

First, in Fig. \ref{skr_compare}, the SKR of our proposed scheme outperforms the others in low SNR regime (SNR$<15$dB). This is attributed to the usages of the largest singular values of Alice's and Bob's received signal matrix. Such a largest singular value is able to suppress the noise component by extracting the main basis from a noisy matrix. This thereby makes the SKR of the proposed scheme insensitive to the changes of the SNR, which guarantees the performance in low SNR region.

Second, when comparing with the pilot based PL-SKG under single Eve, the SKR of our proposed scheme is lower in high SNR regime (SNR$>15$dB). This is because of the usage of the random Gaussian matrix in our proposed scheme. Compared to the globally known pilot sequence, the random matrix induces extra uncertainties when estimating the common channel properties (e.g., the singular value or the RSS). 

Third, in Fig. \ref{skr_compare}, with the increase of the number of colluded Eves, the SKR of the pilot based PL-SKG decreases, and becomes lower than that of our proposed scheme. For example, when $10$ colluded Eves are involved, the SKR of the pilot based method decline to $2$bit/sample. The value is further decreasing to $0$bit/sample when $50$ colluded Eves are placed. This is due to the usage of the globally known pilot sequence for legitimate users' channel estimation, which enables the proposed colluded Eves (in Section II. C) to estimate legitimate channel property and further decode the secret key.
By comparison, our proposed RGM PL-SKG method shows a stable SKR as $3$bit/sample, even if 100 colluded Eves are considered. Such a SKR advantage has two reasons. Firstly, the replacement of the globally known pilot with the random Gaussian matrix is able to prevent the illegal channel estimation and key leakage to Eves. Secondly, the common singular value property (theoretically proved by Theorem \ref{theo2} and Proposition \ref{prop1}) leads to a high mutual information between $\sigma_1^A$ and $\sigma_1^B$, and thereby guarantees the stable SKR performance.

\section{Conclusion}
Physical layer secret key leverages the dynamics and randomness of the reciprocal legitimate channels for key generation, which has been demonstrated as a promising techniques to secure the confidentiality of the wireless communications. For the static or scarcely random environment, the advancement of IRS enables to create artificial randomness by its phase controller, which, combined with the current PL-SKG method, is able to alleviate the key shortage and improve the secrecy performance. However, most of the current researches overlook the fact that the IRS induced random phase is also revealed in the Eves' signals. This enables the Eve nodes to collude and estimate the legitimate channels and further the secret key, via the globally known pilot sequence. 

In this work, to prevent the illegal channel estimation by Eve nodes, we avoided the usage of a global known pilot sequence. Instead, we proposed to leverage on random matrix theory to develop a novel PL-SKG process and created a new idea for secret key design. We proved that, when sending appropriate random Gaussian matrices, the singular values of the Alice's and Bob's received signals follow a similar probability distribution. Leveraging this, we proposed a random Gaussian matrix based PL-SKG, which uses the common singulars of the received random matrices, and avoids the usage of the globally known pilot that can be used by Eves for channel estimation and key decoding. 

Our results demonstrated: (i) comparatively superior (up to 300\%) SKR in the low SNR regime, attributed to the noise resistance ability of the singular values, and (ii) generally improved SKR performance for multiple colluded Eves. We therefore believe our proposed random matrix theory for PL-SKG a promising method, which shows a novel signal processing paradigm to secure the wireless communication channels.

\bibliographystyle{IEEEtran}
\bibliography{main.bib}

\end{document}